\documentclass[aps,floats]{revtex4}
\usepackage{amsmath}
\usepackage{graphicx,epsfig}
\begin{document}
\bibliographystyle {plain}

\def\oppropto{\mathop{\propto}} 
\def\opsimeq{\mathop{\simeq}}
\def\opoverderline{\mathop{\overline}}
\def\operarrow{\mathop{\longrightarrow}}
\def\opsim{\mathop{\sim}}

\def\fig#1#2{\includegraphics[height=#1]{#2}}
\def\figx#1#2{\includegraphics[width=#1]{#2}}

%\newcommand{\fig}[2]{\epsfxsize=#1\epsfbox{#2}} \reversemarginpar 

%%%%%%%%%%%%%%%%%%%%%%%%%%%%%%%%%%%%%%%%%%%%%%%%%%%%%%%%%%%%%%%%%%%%%%%%%%%%
\title{ Directed polymers and interfaces in random media  :  \\
Free-energy optimization via confinement in a wandering tube   } 

 \author{C\'ecile Monthus and Thomas Garel}
 \affiliation{Service de Physique Th\'eorique, 
  Unit\'e de recherche associ\'ee au CNRS, \\
  DSM/CEA Saclay, 91191 Gif-sur-Yvette, France}

%%%%%%%%%%%%%%%%%%%%%%%%%%%%%%%%%%%%%%%%%%%%%%%%%%%%%%%%%%%%%%%%%%%%%%%%%%%%
\begin{abstract}

We analyze, via Imry-Ma scaling arguments, the strong disorder phases that exist in low dimensions at all temperatures for directed polymers and interfaces in random media. For the uncorrelated Gaussian disorder, we obtain that the optimal strategy for the polymer in dimension $1+d$ with $0<d<2$ involves at the same time (i) a confinement in a favorable tube of radius $R_S \sim L^{\nu_S}$ with $\nu_S=1/(4-d)<1/2$ (ii) a superdiffusive behavior $R \sim L^{\nu}$ with $\nu=(3-d)/(4-d)>1/2$ for the wandering of the best favorable tube available. The corresponding free-energy then scales as $F \sim L^{\omega}$ with $\omega=2 \nu-1$ and the left tail of the probability distribution involves a stretched exponential of exponent $\eta= (4-d)/2$. These results generalize the well known exact exponents $\nu=2/3$, $\omega=1/3$ and $\eta=3/2$ in $d=1$, where the subleading transverse length $R_S \sim L^{1/3}$ is known as the typical distance between two replicas in the Bethe Ansatz wave function. We then extend our approach to correlated disorder in transverse directions with exponent $\alpha$
 and/or to manifolds in dimension $D+d=d_{t}$ with $0<D<2$. The strategy of being both confined and superdiffusive is still optimal for decaying correlations ($\alpha<0$), whereas it is not for growing correlations ($\alpha>0$). In particular, for an interface of dimension $(d_t-1)$ in a space of total dimension $5/3<d_t<3$ with random-bond disorder, our approach yields the confinement exponent $\nu_S = (d_t-1)(3-d_t)/( 5d_t-7)$. Finally, we study the exponents in the presence of an algebraic tail $1/V^{1+\mu}$ in the disorder distribution, and obtain various regimes in the $(\mu,d)$ plane.

\end{abstract}

\maketitle

\section{Introduction}

Directed polymers in random media have attracted a lot of interest
for many years, either as interesting disordered models or in relation with
stochastic growth models \cite{revue}.
For polymers in dimension $1+d$ described in the continuum limit
by the partition function
\begin{eqnarray}
 Z= \int {\cal D} \vec r(s) e^{-  \int_0^L ds \left( \frac{d \vec r}{ds} \right)^2
- \beta \int_0^{L} ds V(s,\vec r (s) ) }
\label{partition}
\end{eqnarray}
with an uncorrelated Gaussian random potential
\begin{eqnarray}
 \overline{ V(s,\vec r) V(s',\vec r') }
=   \delta(s-s') \delta^d (\vec r - \vec r')
\label{delta2}
\end{eqnarray}
the phase diagram is the following \cite{revue} : in dimension $d>2$,
there exists a phase transition between a free phase at
   high temperature  \cite{imbrie,cook}, and 
a pinned phase at low temperature :
 this phase transition
has been studied numerically in $d=3$ \cite{golinelli},
exactly on a Cayley tree \cite{dspohn} and on hierarchical lattice
\cite{dgriff}.
 On the contrary, in dimension $d < 2$, there is no free phase,
i.e. any initial disorder drives the polymer into a strong disorder phase.
The marginal dimension $d=2$ has been controversial and deserves
a special discussion \cite{revue}. A strong disorder phase is
 characterized in particular by two exponents $\omega$ and $\nu$ 
 for the free energy $F$ 
and the transverse length scale $R$ 
\begin{eqnarray}
F(L) \sim L^{\omega} \\
R(L) \sim L^{\nu}
\label{defexp}
\end{eqnarray}
with the expected scaling relation $\omega=2 \nu-1$ \cite{scalingrelation}.
In $1+1$, these exponents $\omega$ and $\nu$ are exactly known to be
$\omega=1/3$ and $\nu=2/3$, because in $d=1$, some ``miracles"
happen in various methods : 
via the mapping towards a damped Burgers equation with random forcing,
there exists an exact steady-state distribution that fixes the values
of the exponents \cite{husehenley}; within the replica framework,
there exists exact Bethe Ansatz solutions, that have been studied
either in unbounded
space \cite{kardar,jpho}, or in bounded space \cite{brunet} ; 
there exists an exact combinatorial
solution at zero-temperature \cite{johansson},
as well as other exact results via
the correspondence with stochastic growth models \cite{praehofer}
in the same KPZ universality class. 
So there are plenty of reasons why the exponents are exactly
$1/3$ and $2/3$ in $d=1$. However, various questions
are still open or under debate \cite{revue}, concerning
the generalizations of these exponents in various ways, namely :

(i) in other transverse dimensions $d$

(ii) in the presence of transverse correlations

(iii) for manifolds and interfaces of higher internal dimensions $D$

(iv) for various initial disorder distributions, presenting for instance
algebraic tails.

In this paper, our aim is to present Imry-Ma scaling arguments 
that allow to analyze these various generalizations
in a unified framework, by a proper identification
 of the underlying optimal strategy in each case.
The paper is organized as follows.

In Section \ref{twoarg}, we recall 
 the two `local' Imry-Ma arguments proposed in 
\cite{garelvaria} for the directed polymer in $1+d$ dimensions
in favorable and unfavorable regions.
In Section \ref{global}, we propose a global optimization mechanism
between the energy gained by a confinement in a favorable `tube'
and the global elastic energy to find the best favorable tube available. 
This strategy fixes a confinement exponent, a global wandering exponent,
as well as a free-energy exponent and the form of the left tail for
 the free energy probability distribution.
In Section \ref{dimcorre}, we generalize our approach to 
other correlations in transverse directions
and/or to other internal dimensions : we discuss in particular the cases
of directed polymers with correlated transverse disorder,
of interfaces with random bond disorder
and of interfaces with random field disorder.
In all these cases, we find exponents in agreement with the replica-scaling analysis by Zhang \cite{zhangreplicascaling}, and we thus discuss the dictionary
between our approach via Imry-Ma argument in one disordered sample
and the Zhang analysis in replica space.
Finally in Appendix \ref{sectionmu}, 
we generalize our approach for directed polymers in dimension
$1+d$ in the presence of an algebraic tail 
$1/V^{1+\mu}$ in the disorder distribution, 
and we obtain various regimes in terms of the two parameters $(\mu,d)$.

\section{ Two local Imry-Ma arguments for favorable and unfavorable regions}

\label{twoarg}

In this Section, we recall in details the two Imry-Ma arguments proposed in 
\cite{garelvaria} which constitute the basis of all our discussion.
(In \cite{garelvaria}, these arguments were given to interpret the two
types of solution found via the disorder-dependent variational method).
As in other contexts, the Imry-Ma argument \cite{imryma} 
begins with the evaluation
of the typical energy associated to the disorder in a certain volume.
Here, for a polymer of length $l$ and transverse length $r$, 
the dimensional analysis of the correlator (\ref{delta2})
yields the following scaling for
the typical random energy 
\begin{eqnarray}
 \left( \int_{0}^l ds  V(s,\vec r(s)) \right)_{typ}
\sim \pm u\sqrt{  \frac{  l }{ r^d} } 
\label{imrymaarg}
\end{eqnarray}
where $u$ is a random variable of order 1.
Regions with $u=u_- > 0$ correspond to globally ``unfavorable regions",
whereas regions with $u=-u_+ < 0$ correspond to globally 
``favorable regions".

In other contexts, such as random-field Ising models \cite{imryma},
the Imry-Ma argument then consists in comparing 
the energy cost in creating domain walls with 
the typical energy gained by taking advantage of the favorable fluctuations
of the disorder. Here in the polymer context, the energy 
coming from the disorder has to be compared with 
the entropy cost, that can take two different forms \cite{garelvaria} :
for a swollen polymer $ r \gg l^{1/2}$, 
the entropy cost consists in an elastic term $T r^2/l$ ,
whereas for a confined polymer $r \ll l^{1/2}$,
the entropic cost consists in a confinement term $T l/r^2$.
These two possibilities lead to two different Imry-Ma arguments,
that can be associated to unfavorable and favorable regions
\cite{garelvaria} as we now explain. 

\subsection{ Imry-Ma argument with the elastic term for ``unfavorable regions" }

The free-energy of an unfavorable region of length $l_-$ and
transverse length $r_-$ is the sum of 
the elastic term 
$T r_-^2/l_-$, that represents an entropic cost,
 and the energy cost $u_-\sqrt{  \frac{  l_- }{ r^d_-} }$
from the unfavorable fluctuation of the disorder (\ref{imrymaarg}) 
\begin{eqnarray}
f_-  \sim T \frac{r_-^2}{l_-} + u_-\sqrt{  \frac{  l_- }{ r^d_-} }
\label{defE-}
\end{eqnarray}
The minimization with respect to $r_-$ yields, after dropping
numerical prefactors 
\begin{eqnarray}
r_- \sim  \left( \frac{u_-}{T} \right)^{\frac{2\nu_-}{3}} l_-^{\nu_-} \ \ \rm{with} \ \  \nu_- = \frac{3}{4+d}
\label{R-versusL-}
\end{eqnarray}
in particular $\nu_-(d=1)=3/5$.
The corresponding scaling for the free energy (\ref{defE-})
of this unfavorable region reads
\begin{eqnarray}
f_- \sim T^{\frac{d}{4+d}} u_-^{\frac{4}{4+d} } l_-^{\omega_-} \ \ \rm{with} \ \    \omega_-  
= 2 \nu_- -1 = \frac{2-d}{4+d}
\label{E-versusL-}
\end{eqnarray}
in particular $\omega_-(d=1)=1/5$.
These exponents $\nu_-$ and $\omega_-$ 
actually correspond to the direct dimensional analysis
of the initial Hamiltonian, and are usually called
 ``Imry-Ma exponents" or ``Flory exponents" in the more general context of 
interfaces and manifolds in random media
\cite{grinstein, fisherrfim,kardarmani,nattermann,thhmani,zhangreplicascaling,soluRSB,henrirfim}.
In dimension $d=1$, these exponents $(\nu_-=3/5,\omega_-=1/5)$ are also 
the exponents predicted for the full polymer  
by the Replica Gaussian variational
Ansatz with replica symmetry breaking \cite{soluRSB},
in contrast with the correct exponents $(\nu=2/3,\omega=1/3)$
for the full polymer
found by the replica symmetric Bethe Ansatz solution \cite{kardar}.

Here, we stress that the above Imry-Ma dimensional analysis should not a priori
be applied blindly to the full polymer, but only to the
unfavorable regions. Our conclusion for the moment being
is thus the following :
if the polymer has to cross an unfavorable region  $u_-<0$,
it will behaves as follows
when the dimension $d$ varies :
 
(i) for $0<d<2$, the polymer will adopt a wandering exponent $\nu_- = \frac{3}{4+d}>1/2$,
and the free energy will have for exponent
$\omega_-  = \frac{2-d}{4+d}>0$.

(ii) for the marginal case $d=2$, the wandering exponent 
reaches the free value $\nu_-(d=2)=1/2$
and the free energy exponent vanishes $\omega_- (d=2) = 0$.
A more refined analysis thus becomes necessary.

(iii) for $d>2$ the above Imry-Ma argument that would yield
$\nu_-<1/2$ breaks down,
since for a confined polymer, 
the elastic free energy $r_-^2/l_-$
has to be replaced by the confinement free energy $l_-/r_-^2$.
However, in this case, the free energy is minimum in the limit $r_- \to \infty$,
that does not correspond to a confined configuration. 
So at the level of this scaling analysis,
the only consistent possibility for $d>2$ is that the polymer will keep its 
free exponent $\nu_-=1/2$ corresponding to a finite elastic free energy,
and the disorder potential then corresponds to a subleading
term of order $l^{(2-d)/2}$, which is what happens
in the high temperature phase.

\subsection{ Imry-Ma argument with the confinement term for ``favorable regions"  }
 
The free energy of a favorable region
$(l_+,r_+)$  is the balance between the confinement term $T l_+/r_+^2$,
representing the the entropy loss
due to the confinement,  
and the energy gain $u_+\sqrt{  \frac{  l_+ }{ r^d_+} }$
from a favorable fluctuation of the disorder (\ref{imrymaarg})
\begin{eqnarray}
f_+  \sim T \frac{l_+}{r_+^2} - u_+\sqrt{  \frac{  l_+ }{ r^d_+} }
\label{defE+}
\end{eqnarray}
The minimization with respect to $r_+$ yields, after dropping
numerical prefactors,
\begin{eqnarray}
r_+ \sim   \left( \frac{T}{u_+} \right)^{2 \nu_+} l_+^{\nu_+} \ \ \rm{with} \ \  \nu_+ = \frac{1}{4-d}
\label{R+versusL+}
\end{eqnarray}
in particular $\nu_+(d=1)=1/3$.
The corresponding scaling for the total free energy of this favorable region reads
\begin{eqnarray}
f_+ \sim -  T^{-d \nu_+} u_+^{4 \nu_+} l_+^{\omega_+} \ \ \rm{with} \ \    \omega_+  
= 1- 2 \nu_+ =\frac{2-d}{4-d}
\label{E+versusL+}
\end{eqnarray}
in particular $\omega_+(d=1)=1/3$.
In contrast with the exponents $(\nu_-,\omega_-)$ coming
from a direct dimensional analysis of the Hamiltonian,
the exponents $(\nu_+,\omega_+)$ take into account 
the physical idea that it can be better for the polymer
to remain confined in a region to take advantage of favorable fluctuations
of the disorder. To our knowledge, these exponents $(\nu_+,\omega_+)$
have not been considered previously, except in \cite{garelvaria}
where they have been introduced.

Our conclusion for the moment being
is thus the following : 
if the polymer has to cross a favorable region $u_+>0$,
it will behaves as follows
when the dimension $d$ varies :

(i) for $0<d<2$, the polymer will adopt a confinement exponent $\nu_+ = 
\frac{1}{4-d}<1/2$,
and the free energy will have for exponent
$\omega_+  = \frac{2-d}{4-d}>0$.

(ii) for the marginal case $d=2$, the confinement exponent reaches the free value $\nu_+(d=2)=1/2$
and the free energy exponent vanishes $\omega_+ (d=2) = 0$.
A more refined analysis thus becomes necessary.

(iii) for $d>2$ the above Imry-Ma argument that would yield
$\nu_+>1/2$ breaks down, since the polymer is not confined anymore.
If one replaces the confinement term $l_+/r_+^2$
by the elastic term $r_+^2/l_+$, the total free energy will be minimum
for $r_+ \to 0$ corresponding to a confined configuration.
So at the level of this scaling analysis, exactly as
in unfavorable regions,
the only consistent possibility for $d>2$ is that the polymer will keep its 
free exponent $\nu_+=1/2$ corresponding to a finite elastic free energy,
and the disorder potential then corresponds to a subleading
term of order $l^{(2-d)/2}$, which is what happens
in the high temperature phase.

\subsection{ Discussion}

In this Section, we have described via Imry-Ma arguments
what typical scalings should be expected from a polymer that is obliged
to go through a given favorable region or through
a given unfavorable region.
In dimension $0<d<2$, this analysis yields two sets of non-trivial exponents $(\nu_-,\omega_-)$ and $(\nu_+,\omega_+)$ for the two types of regions,
whereas for $d > 2$, the only self-consistent
exponents in the above Imry-Ma scaling analysis
are the exponents of the high-temperature phase.
This suggests that the pinned phase existing in dimension $d >2$ 
at low temperature is very different in nature 
from the physics in dimension $d<2$
and requires a different type of analysis.
In the following, we will thus only consider the cases $0<d<2$,
where the disorder is strong at all scales and changes the free exponents
both in favorable and in unfavorable regions.

\section{ Structure of the full polymer of length $L$ 
in dimension $0<d<2$ }

\label{global}

In this Section, we consider the standard situation
of a polymer of length $L$ whose origin is fixed
and whose end-point is free. We discuss what is the best strategy
to obtain a minimum free energy, in terms of the favorable regions
described in previous Section.

\subsection{ Global optimization on scale $L$ and exponents}

The simplest strategy that seems optimal at large scale $L$
is the following :
the polymer will try to find a favorable region
of length $L_+ \sim L$, of transverse length $R_+ \sim L^{\nu_+}$
and of free energy $F_+ \sim - L^{\omega_+}$.
The only degree of freedom available to find this very favorable
region is the global orientation $R_G \sim L^{\nu}$, 
with respect to the horizontal line, of the tube of radius $R_+$
starting from the origin forming the favorable region.
To find the best favorable region available, the polymer
can afford a global elastic cost $T R_G^2/L$ that is
 at most of the same order of magnitude
of the free energy $F_+ \sim L^{\omega_+} $ of the favorable region
it is looking for. The balance between these two terms 
yields the following global transverse distance 
\begin{eqnarray}
R_G \sim L^{\nu} \ \ \rm{with} \ \ \nu=\frac{1+\omega_+}{2}=\frac{3-d}{4-d}
\label{resnu}
\end{eqnarray}
and the corresponding free energy
\begin{eqnarray}
F \sim - L^{\omega} \ \ \rm{with} \ \ \omega=\omega_+=\frac{2-d}{4-d}
\label{resomega}
\end{eqnarray}
These two exponents are thus the generalizations in dimension $0<d<2$
of the well known exact exponents $\nu=2/3$ and $\omega=1/3$ in $d=1$ 
\cite{husehenley,revue}. 
Moreover, our description also yields the subleading transverse length scale
\begin{eqnarray}
R_+ \sim L^{\nu_+} \ \ \rm{with} \ \ \nu_+=\frac{1}{4-d}
\label{rplus}
\end{eqnarray}
representing the radius of the ``tube" of the favorable region,
that generalizes the transverse length scale $L^{1/3}$ 
introduced in \cite{zhang} to characterize the size of a ``family",
i.e. paths having free energy differences of order $O(1)$.
This subleading transverse length scale $L^{1/3}$
was also interpreted in \cite{revue} as the typical distance
between two replicas in the Bethe Ansatz replica wavefunction \cite{kardar},
whereas the scale $L^{2/3}$ represent the displacement of all replicas as a whole. 

The physical meaning of the present Imry-Ma scaling analysis
is thus the following : the configuration of the polymer
is determined by a global optimization mechanism at 
the largest scale :
the polymer chooses the best ``tube" of radius $R_+\sim L^{\nu_+}$
among all tubes available labeled by the global orientation
$\rho=R_L/L^{\nu}$ defined by the transverse distance $R_L$
of the end point. The number of different tubes available for the choice
thus scales as
\begin{eqnarray}
N \sim \frac{R^d}{R_+^d} \sim L^{\gamma}
\ \ {\rm with} \ \ \gamma=d (\nu-\nu_+)=\frac{ d (2-d)}{ 4-d}
\label{ntubes}
\end{eqnarray}
in particular $\gamma(d=1)=1/3$.
This number is large enough to find a `good' tube in a arbitrary sample.

\subsection{ Tail of the distribution of the free energy }

The Imry-Ma analysis for favorable regions can be used 
to study the tail of the probability distribution of the rescaled free energy.
Indeed, the asymptotic behavior of the random variable $u_+$
in the Imry-Ma argument (\ref{imrymaarg},\ref{defE+})
is expected to follow to the Gaussian distribution 
\begin{eqnarray}
P(u_+) \opsimeq_{u_+ \to \infty} e^{- u_+^2}
\label{gauss}
\end{eqnarray}
The same idea has been already used in the context of the random field
XY model \cite{garelgauss}, and
in the context of a disordered heteropolymer ,
where it was shown to be in full agreement with a disorder dependent
real-space RG analysis (see the Appendix of \cite{c_polyrsrg}).
In the present context, we stress that the Gaussian tail (\ref{gauss}) 
is valid for an initial Gaussian disorder,
whereas the presence of an algebraic tail of arbitrary order
in the initial disorder will generate a different tail,
as discussed in Appendix \ref{sectionmu}. 

Here, for a Gaussian initial disorder,
the Gaussian tail (\ref{gauss}) yields,
via the change of variables (\ref{E+versusL+}), 
the following decay for the probability distribution
of the rescaled free energy $f_+=F_+/L_+^{\omega_+}$ 
\begin{eqnarray}
{\cal P}_+(f_+) \oppropto_{f_+ \to -\infty} 
T^{d/4} \left( \vert f_+ \vert \right)^{\frac{\eta_+}{2}-1}
e^{- T^{d/2} \vert f_+ \vert^{\eta_+} }
\ \ \ {\rm with } \ \ \eta_+= \frac{1}{2 \nu_+} =\frac{4-d}{2}
\label{taile+}
\end{eqnarray}

At the level of exponents where we work, since $L_+ \sim L$,
the tail of the probability distribution ${\cal P}(f)$ for the rescaled free energy 
$f=F/L^{\omega}$ of a polymer of length $L$ will
thus be given by the same form (\ref{taile+}) 
 \begin{eqnarray}
{\cal P}(f) \opsim_{f \to -\infty} {\cal P}_+(f)
\end{eqnarray}
In particular, the exponent in the exponential in one dimension
is $\eta_+(d=1)=3/2$,
a value that agrees with
the replica-scaling predictions \cite{kardar,zhangreplicascaling}
and with the numerical simulations \cite{kim}.

However, to be fully consistent with the approach we propose,
we should take into account that the polymer actually
chooses the best tube among of large number $N$ (\ref{ntubes})
of independent tubes. In this interpretation, the random variable $u_+$
is not just a random variable drawn with a distribution
having the Gaussian tail (\ref{gauss}), but it is the maximal
value $u_{max}$ drawn among $N$ independent variables, i.e. its
distribution reads
 \begin{eqnarray}
\rho_N(u_{max}) = N P(u_{max}) \left[ 1- \int_{u_{max}}^{+\infty} du P(u) \right]^{N-1} \opsim_{N \to \infty} N e^{- u_{max}^2}
e^{- N \int_{u_{max}}^{+\infty} du e^{- u^2} }
\label{distrimaxg}
\end{eqnarray}
In particular, $u_{max}$ grows logarithmically in
 $N$ and thus in $L$ (\ref{ntubes})
 \begin{eqnarray}
u_{max} \sim \sqrt{ \ln N} \sim \sqrt{ \ln L }
\label{umaxgauss}
\end{eqnarray}
This would lead (\ref{E+versusL+}) to a logarithmic correction to the exponent
for the free-energy (\ref{resomega}), i.e.
\begin{eqnarray}
F \sim - L^{\omega} u_{max}^{4 \nu_+} \sim -L^{\omega} 
(\ln L)^{\frac{1}{\eta_+}}
\label{resomegaln}
\end{eqnarray}

Is this logarithmic correction a reality or an `artefact' of our interpretation?
On one hand, the comparison with the results of most other studies
on the subject
suggests that this logarithmic correction is spurious.
On the other hand, within our approach, it is not clear to us
why this logarithmic correction should be disregarded.
In particular, if the initial disorder distribution is not Gaussian
 but presents an algebraic tail of index
$(1+\mu)$, it is precisely this mechanism of choosing
 the best variable $u_+$
that opens the possibility of obtaining different global exponents
even if the variance is finite $\mu>2$,
as discussed in Appendix \ref{sectionmu}.
We note moreover that the presence of some logarithmic factors coming from
extremal statistics has already been proposed and numerically studied
for the directed polymer in $1+1$ \cite{alavaextremal},
as well as in another context involving polymers
in random media \cite{catesball}.
In conclusion, even if the exact solution at zero temperature \cite{johansson}
has no logarithmic correction, since we are not aware of
exact solutions at non-zero temperature, and since the limit of
zero-temperature cannot be discussed within our approach
(see the discussion below in \ref{zerotemp}), it seems that
the presence of this logarithmic correction for the free energy at non-zero temperature is a possibility that cannot be completely ruled out,
at least to the best of our present knowledge.
If there is a proof in the future that there is no logarithmic correction,
this would probably mean that the `best' tube is not simply
the tube having the maximal rescaled variable $u$, but the tube
having the best structure on smaller scales than the global scale.

\subsection{ Remarks on the zero-temperature limit }

\label{zerotemp}

In this paper, we have considered that the temperature appears only
in front of the random potential in the partition function (\ref{partition})
and not in front of the Wiener measure for the Brownian paths.
In this case, the elastic term is an entropy coming from
the probability $e^{-R^2/L}$ to be at distance $R$ in time $L$
for a Brownian motion. In particular, 
this elastic term is not present at $T=0$, where the problem on the 
hypercubic lattice in $(1+d)$ consists in finding the best path
among the $(1+d)^L$ possible paths, i.e.
 on the lattice, a path is either allowed or forbidden, 
there is no elastic energy at $T=0$.
This corresponds to the usual model for numerical simulations 
on directed polymers at $T=0$.
However, many authors are also interested by the models
where the elastic term $R^2/L$ is an energy,
i.e. in the partition function (\ref{partition}), the temperature appears
not only in front of the random potential but 
as a global factor in front of the two terms in the exponential 
(\ref{partition}). In this context, the problem at $T=0$ 
consists in finding the best path that optimizes the sum
of the elastic energy and the random energy.

However, in both cases, within our approach, 
the limit of zero-temperature
$T \to 0$ is very singular,
because the entropy due to confinement that played a crucial role at
$T \neq 0$ disappears at $T=0$.
Let us briefly see what happens as $T \to 0$ in the
Imry-Ma arguments energy/entropy that we have discussed.
For unfavorable regions, the prefactor of the
transverse scale $r_-$ (\ref{R-versusL-})
diverges and the prefactor of the free-energy (\ref{E-versusL-}) vanishes :
indeed, at $T=0$, the free-energy (\ref{defE-}) only contains
the random energy from the disorder (and no entropic elastic term anymore) :
its minimization corresponds to a transverse scale $r_-$ as big as possible,
i.e. on a lattice $r_- \sim l_-$
that leads to an energy $e_- \sim u_- l_-^{(1-d)/2}$. 
For favorable regions,  
 the prefactor of the
confinement scale $r_+$ (\ref{R+versusL+})
vanishes, whereas the prefactor of the free-energy (\ref{E+versusL+}) 
diverges :
indeed, at $T=0$, the free-energy (\ref{defE+}) only contains
the random energy from the disorder 
(and no entropic confinement cost anymore) :
its minimization corresponds to a confinement
scale $r_+$ as small as possible, i.e. on the lattice $r_+ \sim 1$
corresponding to a unique path,
that leads to an energy of order $e_+ \sim -u_+ l_+^{1/2}$.

In conclusion, our description with Imry-Ma arguments
cannot be used to understand the zero-temperature limit,
even if there are very direct relations between $T \neq 0$ and $T=0$ :
in $d=1$, the wandering exponent of the best tube at finite temperature
$\nu=2/3$ is the same as wandering exponent 
of the best path at zero temperature, and 
the subleading transverse scale $R_+ \sim L^{1/3}$ that 
represents in our approach the confinement scale at non-zero temperature,
had been previously identified in zero-temperature numerical simulations
as the typical scale for the first excited states of finite energy above the ground state \cite{zhang,perlsman}.

\subsection{ What is the sub-structure of the polymer at smaller scales ?  }

From the point of view of the Imry-Ma scaling analysis
proposed in this paper, it is clear that
the global exponents $\omega$ (\ref{resomega}) and $\nu$ (\ref{resnu})
are completely constrained by a global optimization mechanism
at the biggest scale. However, once the best global tube has been chosen,
it seems natural to expect that
the polymer has a ``cascade" of optimizations to make
on smaller and smaller scales within the large scale constraints.
In particular, we expect that the extensive contribution to the free-energy
comes from the small scales, since the polymer has 
to gain a finite contribution at each step on average.
However, we also expect that the polymer cannot 
avoid frustration on all scales, i.e. it will be obliged
sometimes to cross ``unfavorable regions" characterized by the exponents
 $\nu_-$ and $\omega_-$ (\ref{R-versusL-},\ref{E-versusL-}).
A first interesting question is : what is the scale $L_-$
of the biggest unfavorable region inside the globally favorable tube?
A more general question concerns the hierarchical  
organization on smaller and smaller scales.
Indeed, the ultrametric tree structure of local optimal paths 
\cite{kardarzhang,perlsman} as well as Monte-Carlo simulations at finite temperature
\cite{yoshino} suggest that they are favorable and unfavorable
regions on various scales.
In conclusion, it would be very interesting to have a statistical description 
of the substructure of the global tube, but this goes beyond the present work.

\subsection{ Discussion of some consequences}

The idea of favorables tubes 
where the polymer remains confined at finite temperature,
with a confinement radius $L^{\nu_+}$ much smaller
than the wandering scale $L^{\nu}$,
gives a more precise meaning to the notion of `states' developed
in the droplet scaling theory \cite{fisherhuse,fisherhwa}.
It is thus interesting to mention briefly some important consequences
in dimension $0<d<2$ with the values of the global exponents that
we have obtained.
(It would be of course very interesting to give a more precise 
characterization of the `states' at low temperature in dimension $d \geq 2$,
but this is left for future work).

\subsubsection{ Statistics of the effective random potential
 for the end-point  }

The effective Hamiltonian seen by the free end-point $\vec r =\vec r(L)$ 
can be decomposed into \cite{fisherhuse,fisherhwa}
\begin{eqnarray}
H_{eff}= T \frac{\vec r^2}{ 2L} +L^{\omega} \Phi \left( \vec \rho =\frac{\vec r}{ L^{\nu} } \right)
\label{defheff}
\end{eqnarray}
where the first term represents the elastic free energy
and the second term an effective random potential that has been rescaled
with the global exponents, and
whose statistics has to be elucidated.
The rescaled effective potential $\Phi (\rho)$, 
has been exactly determined in $d=1$ \cite{praehofer} : it is
an `Airy process' \cite{praehofer}
that behaves locally as a random walk  $ \sqrt{\rho}$ as $\rho \to 0$
and that saturates towards a constant at large distances $\rho \to \infty$,
in agreement with the previous numerical study \cite{mezard}.
 More generally, in dimension $d$, the rescaled effective potential 
is expected to be independent of $L$ at short distance $\rho \to 0$ \cite{fisherhuse},
and thus the exponent $\sigma$
defining the power-law behavior of the effective potential
at short distances 
\begin{eqnarray}
 \Phi (\vec \rho ) \oppropto_{\rho \to 0} \vert \vec \rho \vert^{\sigma}
\end{eqnarray}
is not a new exponent, but is a function 
of the two basic exponents \cite{fisherhuse,fisherhwa}
\begin{eqnarray}
\sigma= \frac{\omega}{\nu}
\label{expsigma}
\end{eqnarray}
With the values obtained before (\ref{resnu},\ref{resomega}),
we thus obtain in dimension $0<d<2$
\begin{eqnarray}
\sigma= \frac{2-d}{3-d}
\end{eqnarray}
that generalizes the random walk behavior $\sigma=1/2$ of the $d=1$ case.
The physical meaning of the relation (\ref{expsigma}) is the following 
\cite{kardarzhang,fisherhuse,fisherhwa} :
optimal configurations whose end-points are separated by a 
 distance $r$ typically merge at a distance $l \sim r^{1/\nu}$
and are then identical up to the origin, so that
their difference in free energy then scales as $l^{\omega} \sim  r^{\omega/\nu}$.
For large distance $r \gg L^{\nu}$ however, the two paths meet
only the origin, and thus experience statistically independent disorders.
This is why the random potential saturates for large separation
\cite{fisherhuse,fisherhwa}.

\subsubsection{  Large-scale thermal fluctuations  }

In the droplet scaling theory \cite{fisherhuse,fisherhwa}, 
there are large-scale thermal fluctuations that come from 
the rare nearly degenerate `states'. In our description with
favorables tubes, if we consider the excitations that involve
a length $l$ of the polymer : there exists 
a large number $N \sim l^{\gamma}$ (\ref{ntubes}) of
other less favorable tubes of energies $l^{\omega}$ 
with transverse distance $\Delta \sim l^{\nu}$ with respect to the best tube.
As a consequence, with probability $T/l^{\omega}$,
there exists an excited tube with a free-energy difference of order
$T$ with respect to the best tube \cite{fisherhuse}. 
The power-law distribution $P(\Delta)$ 
of the end-point transverse scale $\Delta$
of thermal excitations computed via dynamical field theory \cite{fisherhwa}
can be understood as follows 
\begin{eqnarray}
P(\Delta) \Delta^{d-1} d \Delta = \frac{T}{l^{\omega}} \theta(l<L)
\end{eqnarray}
 with the corresponding size $l \sim \Delta^{1/\nu}$, so that
\begin{eqnarray}
P(\Delta) \sim \frac{1}{\Delta^{b} }   w \left( \frac{\Delta}{L^{\nu}} \right)
\ \ {\rm with }  \   b=d+ \frac{\omega}{\nu} = d+2-\frac{1}{\nu}
\end{eqnarray}
where the cut-off $w$ comes from the finite-size of the polymer $l \leq L$.
With the values obtained before (\ref{resnu},\ref{resomega}),
we thus obtain in dimension $0<d<2$, the power-law exponent
\begin{eqnarray}
b= \frac{2+2 d -d^2}{3-d}
\end{eqnarray}
that generalizes the exponent $b(d=1)=3/2$.

\subsubsection{ Chaos exponent  }

The sensitivity to small changes in the disorder distribution
\cite{zhang,feigelman,shapir}
or to small changes in temperature \cite{fisherhuse}
have been interpreted at a scaling level as follows :
if the small random perturbation $\epsilon$
induces a change of order $\epsilon L^{\alpha}$ for the free-energy
of the previous state, there is an instability if $\alpha>\omega$
for lengths larger than $L_c(\epsilon) \sim \epsilon^{- c}$ 
with the so called chaos exponent $c=\frac{1}{\alpha-\omega}$.
For a random bond perturbation \cite{zhang,feigelman,shapir}, 
and for a temperature change \cite{fisherhuse}, the exponent 
of the perturbation is in both cases $\alpha=1/2$. 
For the temperature change, this comes from the behavior of entropy
fluctuations $\Delta S \sim L^{1/2}$, that was conjectured in
\cite{fisherhuse} and numerically checked \cite{fisherhuse,entropie}.
With the values obtained before (\ref{resomega}),
the chaos
exponent thus reads in dimension $0<d<2$ 
\begin{eqnarray}
c= \frac{1}{1/2-\omega} = \frac{d}{2 (4-d) }
\end{eqnarray}
that generalizes the well-known exponent $c(d=1)=1/6$.
Finally, let us mention that
more general perturbations with other exponents $\alpha$ have also been studied
in \cite{shapir,sales}, 
and that the scaling form for the free-energy decorrelation
has been recently computed on a Berker lattice \cite{bouchaud}.

\section{ Generalization to other dimensions and/or other transverse correlations}

\label{dimcorre}

In this section, we generalize our approach to the case of a D-dimensional 
manifold embedded in a space of total dimension $D+d=d_t$
in the solid-on-solid approximation : the manifold is described
by a d-dimensional vector field $r(x)$, where $x$ is the D-dimensional
vector of internal coordinates.
We consider the standard Hamiltonian
\begin{eqnarray}
H =  \int d^D x \sum_{\mu=1}^D \left( \frac{\partial r}{\partial x_{\mu}} \right)^2
+ \int d^D x V(x,r(x) ) 
\end{eqnarray}
where $V(x,r)$ is a Gaussian random potential with correlator
\begin{eqnarray}
 \overline{ V(x,r) V(x',r') }
=   \delta^D(x-x') C ( \vert r-r' \vert)
\label{corregene}
\end{eqnarray}
where the asymptotic behavior of the correlation in transverse directions
is governed by some exponent $\alpha$
\begin{eqnarray}
C( r ) \oppropto_{r \to \infty} r^{\alpha}
\label{defalpha}
\end{eqnarray}
For instance, the case of local disorder characterized by 
delta correlations also in transverse directions corresponds
via scaling to the case $\alpha=-d$, whereas a random-field disorder corresponds to $\alpha=1$. 

The dimensional analysis of the elastic term $L^{D-2} R^2$ shows that 
the free behavior in the absence of disorder is characterized
by the exponent 
\begin{eqnarray}
\nu_{free}(D)=\frac{2-D}{2}
\end{eqnarray}
that generalizes the random walk exponent $\nu_{free}(D=1)=1/2$ of the polymer case. In the following, we will only consider the cases $0<D<2$
where $\nu_{free}>0$.

\subsection{ Local Imry-Ma argument with the elastic term }

The generalization of the free energy (\ref{defE-}) reads
\begin{eqnarray}
f_-  \sim T \frac{r_-^2}{l_-^{2-D}} + u_- l_-^{D/2} r_-^{\alpha/2}
\label{defE-gene}
\end{eqnarray}

For $\alpha u_-<0$, the minimization of the free energy 
 with respect to $r_-$ yields
\begin{eqnarray}
r_- \sim  u_-^{\frac{2}{4-\alpha}} l_-^{\nu_-} \ \ \rm{with} \ \  \nu_- = \frac{4-D}{4-\alpha}
\label{R-versusL-gene}
\end{eqnarray}
and the corresponding scaling for the free energy reads
\begin{eqnarray}
f_- \sim  u_-^{\frac{4}{4-\alpha}}
 l_-^{\omega_-} \ \ \rm{with} \ \    \omega_- = 2 \nu_- -(2-D) 
= \frac{2D+(2-D) \alpha}{4-\alpha}
\label{E-versusL-gene}
\end{eqnarray}
This solution is consistent as long as $\nu_->\nu_{free}$
i.e. since $0<D<2$
\begin{eqnarray}
  - \frac{2D}{2-D} < \alpha < 4
\label{vali-}
\end{eqnarray}

The condition $\alpha u_-<0$ shows that the physical interpretation
is very different according to the sign of $\alpha$.
For all $\alpha<0$, as in the special case of uncorrelated disorder $\alpha=-d$,
the solution obtained corresponds to $u_->0$, i.e. to a 
unfavorable region, where the global free energy 
is positive $f_->0$.
On the contrary, for $\alpha>0$, the solution corresponds to $u_-<0$
i.e. to a favorable region where the global free energy is negative $f_-<0$.

\subsection{ Local Imry-Ma argument with the confinement term }

The generalization of the Imry-Ma argument (\ref{defE+})
with the confinement entropy reads 
\begin{eqnarray}
f_+  \sim T \left( \frac{l_+}{r_+^{\frac{2}{2-D}}} \right)^D - u_+
l_+^{D/2} r_+^{\alpha/2}
\label{defE+gene}
\end{eqnarray}
(A detailed analysis of the confinement entropy can be found in
\cite{fishercarre}).
For $\alpha u_+<0$, the minimization with respect to $r_+$ yields \begin{eqnarray}
r_+ \sim   u_+^{- \frac{2}{D} \nu_+} l_+^{\nu_+} \ \ \rm{with} 
\ \  \nu_+ = \frac{D(2-D)}{4D+\alpha(2-D)}
\label{R+versusL+gene}
\end{eqnarray}
The corresponding scaling for the total free energy reads
\begin{eqnarray}
f_+ \sim  u_+^{ \frac{4D}{4D +(2-D) \alpha} } l_+^{\omega_+} \ \ \rm{with} \ \    \omega_+  
= D- \frac{2 D}{2-D} \nu_+ =
\frac{D \left[ 2D +(2-D) \alpha \right] }{4D +(2-D) \alpha}
\label{E+versusL+gene}
\end{eqnarray}
This solution is consistent as long as $0<\nu_+<\nu_{free}$,
i.e. 
\begin{eqnarray}
   - \frac{2D}{2-D} < \alpha 
\label{vali+}
\end{eqnarray}

The condition $\alpha u_+<0$ shows that the physical interpretation
is again very different according to the sign of $\alpha$.
For $\alpha<0$, as in the special case of uncorrelated disorder $\alpha=-d$,
the solution obtained corresponds to $u_+>0$, i.e. to a 
favorable region, where the global free energy 
is negative $f_+<0$.
On the contrary, for $\alpha>0$, the solution corresponds to $u_+<0$
i.e. to an unfavorable region where the global free energy is positive $f_+>0$.

\subsection{ Global optimization for $\alpha<0$ }

For $\alpha<0$, as in the polymer case with uncorrelated disorder,
the optimal strategy will be to find a favorable region $L_+ \sim L$
with confinement $R_+\sim L^{\nu_+}$. To find the best region available,
it will be worth to afford a global elastic cost of 
$L^{D-2}R_G^2$ of the same order of magnitude of
the energy gain $E_+ \sim L^{\omega_+}$ of
the favorable region. The balance between these two terms yields
that the global wandering will be
\begin{eqnarray}
R_G \sim L^{\nu} \ \ {\rm with}  \ \ \nu=\frac{\omega_++2-D}{2} =
\frac{ D(4-D) +(2-D) \alpha  }{4D +(2-D) \alpha}
 \end{eqnarray}
The global free energy gain will be
\begin{eqnarray}
F \sim  L^{\omega} \ \ \rm{with} \ \    \omega=\omega_+  
= \frac{D \left[ 2D +(2-D) \alpha \right] }{4D +(2-D) \alpha}
\end{eqnarray}
The subleading transverse length scale characterizing the confinement
reads
\begin{eqnarray}
R_+ \sim L^{\nu_+} \ \ \rm{with} \ \ 
\nu_+ = \frac{D(2-D)}{4D+\alpha(2-D)}
\end{eqnarray}
Finally, the use of (\ref{gauss}) for the random variable $u_+$
yields, via the change of variables (\ref{E+versusL+gene}) to
the following decay for the probability distribution
of the free energy $F \sim F_+$ 
\begin{eqnarray}
{\cal P}(F) \opsimeq_{F \to -\infty} 
\frac{1}{ \vert F \vert }
\left( \frac{\vert F \vert}{L^{\omega}} \right)^{\frac{\eta_+}{2}}
e^{- 
\left( \frac{ \vert F \vert }{L^{\omega}} \right)^{\eta_+  } }
\ \ \ {\rm with } \ \ \eta_+= \frac{4 D + (2-D) \alpha}{2 D}
\end{eqnarray}

\subsection{ Global optimization for $\alpha>0$ }

For $\alpha>0$, the previous picture completely changes, because
the favorable regions are not the confined solutions $(+)$, but the
swollen solutions $(-)$. As a consequence, the global optimization
coincide with a solution $(-)$ : the global exponents 
for the transverse direction and the free energy are thus directly given by
\begin{eqnarray}
\nu && =  \nu_- = \frac{4-D}{4-\alpha} \\
\omega && = \omega_-  
= \frac{2D+(2-D) \alpha}{4-\alpha}
\end{eqnarray}
Finally, the use of (\ref{gauss}) for the random variable $u_-$
yields, via the change of variables (\ref{E-versusL-gene}) to
the following decay for the probability distribution
of the free energy $E \sim E_-$ 
\begin{eqnarray}
{\cal P}(E) \opsimeq_{E \to -\infty} 
\frac{1}{ \vert E \vert }
\left( \frac{\vert E \vert}{L^{\omega}} \right)^{\frac{\eta}{2}}
e^{- 
\left( \frac{ \vert E \vert }{L^{\omega}} \right)^{\eta  } }
\ \ \ {\rm with } \ \ \eta= \frac{4 - \alpha}{2 }
\label{tailecasem}
\end{eqnarray}

We now briefly describe our results with their domain of validity
for the interesting cases of a polymer $D=1$ or of interfaces $d=1$,
before we compare with the results of other methods.

\subsection{ Results for the polymer with decaying correlations 
with exponent $-2<\alpha<0$ }

For a polymer $D=1$ with correlations described
by the exponent $\alpha$ (\ref{defalpha}) with $-2<\alpha<0$, 
the results are the same as for the polymer
with uncorrelated disorder with the replacement $d \to -\alpha$, i.e.
there is confinement with exponent 
\begin{eqnarray}
 \nu_+ && =\frac{1}{4+\alpha}
\end{eqnarray}
with the corresponding exponents for the free energy and the wandering 
\begin{eqnarray}
 \omega && =1-2 \nu_+ =\frac{2 +\alpha}{4+\alpha} \\
\nu && = \frac{3 +\alpha}{4+\alpha}
\label{polyaneg}
\end{eqnarray}
with the tail exponent $\eta_+= 2+\alpha/2$.
The domain of validity in $\alpha$ is limited by $\alpha \to -2$,
where the confinement exponent reaches the free random walk exponent
$\nu_+ \to 1/2$, and by $\alpha \to 0$ where the optimization
mechanism changes.

\subsection { Results for the polymer
 with long-range correlations with exponent $0<\alpha<4$ }

For a polymer $D=1$ with correlations described
by the exponent $\alpha$ (\ref{defalpha}) with $0<\alpha<4$,
there is no confinement, and the exponents are directly given
by the
exponents $(-)$ coming from the direct dimensional analysis
of the Hamiltonian
\begin{eqnarray}
\nu &&    = \frac{3}{4-\alpha} \\
\omega &&   
= 2 \nu-1 = \frac{2+ \alpha}{4-\alpha}
\label{polyapos}
\end{eqnarray}
and the tail exponent $\eta = (4-\alpha)/2$.
The comparison with (\ref{polyaneg}) shows that the
exponents are continuous at $\alpha=0$ with the values
$\nu(\alpha=0)=3/4$ and $\omega(\alpha=0)=1/2$.
Special interesting cases discussed in \cite{revue} are $\alpha=1$,
 and $\alpha=2$. In particular for the latter case $\alpha=2$
that corresponds effectively to the Larkin model, there exists
an exact solution in terms of replicas yielding
the exact free-energy distribution \cite{blatter} :
our approach is in agreement with
the exact results for the global exponents
$\nu=3/2$, $\omega=2$ and for the exponent $\eta=1$ 
of the tail of the free-energy distribution \cite{blatter}.

\subsection{ Results for interfaces with random bond disorder }

For the special cases of interfaces of internal dimension $0<D<2$
in a space of total dimension $D+1=d_t$,
the random bond disorder corresponds to uncorrelated disorder
in the $d=1$ transverse direction, i.e. the effective exponent
is $\alpha=-d=-1<0$. The constraint (\ref{vali+}) leads to the domain of validity $2/3<D<2$, i.e. in terms of the total dimension $d_t$ of space
\begin{eqnarray}
  \frac{5}{3}< d_t < 3 
\label{valiRB}
\end{eqnarray}
The confinement exponent reads 
\begin{eqnarray}
\nu_+ = \frac{(d_t-1)(3-d_t)}{ 5d_t-7}
\end{eqnarray}
Note that it vanishes in the limit $d_t \to 3$, 
because the free exponent $\nu_{free}$
also vanishes in this limit of a two-dimensional surface $D \to 2$
 and is replaced by logarithms, 
the physical meaning being that the confinement
is not a big constraint anymore \cite{fishercarre}.
The corresponding exponents for the free-energy and wandering reads
\begin{eqnarray}
   \omega  
&& = \frac{ (d_t-1) ( 3d_t -5 ) }{5d_t -7} \\
 \nu && =
\frac{ 7 d_t -8 -d_t^2   }{5d_t -7}
\end{eqnarray}
with the tail exponent  $ \eta_+= (5 d_t -7 )/(2 (d_t-1))$.

\subsection{ Results for interfaces with random field disorder }

For the special cases of interfaces of dimension $0<D<2$
in a space of total dimension $D+1=d_t$,
the random field disorder corresponds to the exponent
is $\alpha=+1>0$. The constraint (\ref{vali-}) does not modify
the original constraints $0<D<2$,
i.e. in terms of the total dimension $d_t$ of space
\begin{eqnarray}
  1<d_t<3
\end{eqnarray}
In this case, there is no confinement, and the exponents are given by the
exponents $(-)$ , i.e. they reduce to the well known exponents
coming from the direct dimensional analysis
of the Hamiltonian
\begin{eqnarray}
\nu &&  =  \frac{5- d_t }{3} \\
\omega &&   
= \frac{1+d_t }{3}
\end{eqnarray}
with the left tail exponent $\eta=3/2$, that was also found in \cite{henrirfim}.

\subsection{ Agreement with Zhang replica scaling analysis }

The exponents presented in this Section are in agreement with Zhang replica-scaling analysis \cite{zhangreplicascaling}.
It is thus useful to describe briefly the `dictionary' between the two methods.
Whereas the traditional approach with replicas consists in considering
the limit $n \to 0$ with the possibility of Replica Symmetry Breaking, 
the replica-scaling analysis proposed by Zhang \cite{zhangreplicascaling}
consists in analyzing the replicated problem  within a symmetric treatment
of the $n$ replicas with $n$ large
(for instance $n(n-1)$ is replaced by $n^2$).
The leading orders in $(L,n)$ of the moments are then interpreted
as coming from the tail of the probability distribution of the free-energy.
In Zhang analysis, the case $\alpha<0$ 
 corresponds to an attraction between replicas and leads to a bound state, whereas the case $\alpha>0$ 
 corresponds to a repulsion between replicas with no bound state.
So the presence of a bound state for replicas 
corresponds in our approach to the presence of a confinement tube.
In conclusion, for those who prefer to think in real space
 than in replica space, our approach provides an equivalent
self-contained description in real space; and for those who prefer
to think in replica space, we hope that the translation in real space
is also interesting, and can perhaps be useful in other disordered models.

\subsection{ Comparison with other methods  }

Within the field theoretical framework, the correctness
of the direct dimensional Imry-Ma analysis 
for the random-field disorder $\alpha=1$
was supported by a functional RG analysis \cite{fisherrfim}, because the large distance behavior of the correlations are not renormalized, whereas for random-bond disorder $\alpha=-1$, there are non-trivial renormalization of the exponents \cite{fisherrfim}.
So there should be a critical value $\alpha_c$ 
below which the direct dimensional analysis
does not give the correct exponents.
It has been argued in various RG methods \cite{kardarmani,nattermann,thhmani,medina} 
that (i) below the critical value $\alpha_c$, the wandering exponent $\nu$ sticks to the random bond value $\nu_{RB}$ and would thus be `superuniversal'
(ii) the critical value $\alpha_c$ is strictly negative.
However, since the argument by Fisher \cite{fisherrfim}
that the long-range correlations of the disorder are not renormalized,
naturally stops to apply at $\alpha=0$, it seems to us
that from this point of view, the simplest scenario 
is that the true critical value is actually $\alpha_c=0$, 
in agreement with the prediction of the replica-scaling analysis  
\cite{zhangreplicascaling}
and with our analysis with confinement.
Moreover, we believe that our description involving a confinement mechanism
suggests that a more appropriate
 field theoretical description in the case of transverse decaying correlations
$\alpha<0$ should perhaps include the presence of two different
important transverse length scales, 
and should perhaps involve instantons calculations or other non-perturbative tools to describe the confinement.

Finally, we should mention that the various alternative predictions
\cite{zhangreplicascaling,medina,family} for the exponents in the domain $\alpha<0$ have been tested
 via numerical simulations on kinetic roughening
with various conclusions \cite{margolina,penghavlin}.
We believe that our approach, that explains the meaning
of Zhang replica-scaling exponents in physical space 
strongly suggests that these exponents are indeed the correct ones.

\section{Conclusion}

In this article, we have analyzed via Imry-Ma energy/entropy arguments
the strong disorder phases that exist in low
dimensions at all non-zero temperatures for directed polymers
 and interfaces in random media. 
Within the field of disordered systems, the originality of random manifolds
 is that they have some freedom to `choose'
the disorder variables they see in a given sample, in contrast with 
spin systems for instance that cannot avoid any disorder variables.
Our main result is that they can use this freedom
to follow two qualitatively different strategies : 
(i) for disorder with decaying transverse correlations $\alpha<0$,
the optimal strategy consists in being confined in a wandering tube,
i.e. there are two different transverse scales that are important,
namely the wandering scale and the confinement scale;
(ii) for disorder with long-range transverse correlations $\alpha>0$,
the optimal strategy is to be swollen, i.e. there is only one
important transverse scale, and the exponents are given by a simple
dimensional analysis of the original Hamiltonian.
For the general case of a manifold of internal dimension $D$
in a space of total dimension $D+d$ in the presence of transverse
correlations of exponent $\alpha$, our results for exponents 
agree with Zhang replica-scaling analysis
\cite{zhangreplicascaling}. Our method thus describes the same physics,
but in real space instead of being in replica space,
and this allows
to make the link with the scaling droplet approach :
the `states' at finite temperatures of the scaling droplet theory 
\cite{fisherhuse,fisherhwa}
are interpreted in our approach as Imry-Ma favorable tubes,
where the polymer remains confined,
and it is this confinement in an Imry-Ma favourable tube
that translates into a bound state for replicas
in the replica-scaling analysis \cite{zhangreplicascaling}.

 In conclusion, we hope that our approach via simple scaling arguments
in real space gives a clear insight into the strong disorder phases
in low dimensions, and a complementary point of view with respect to
the other methods.

\section*{Acknowledgments}

It is a pleasure to thank Bernard Derrida for useful comments.

\appendix

\section{ Initial disorder distributions with algebraic tails }
  
\label{sectionmu}

For the lattice uncorrelated models, where each link has an energy $V(i,x)$
drawn with some symmetric law $P(V)=P(-V)$,
the question of the universality with respect to the form of the initial
probability distribution $P(V)$ has attracted a lot of
interest \cite{revue}, because
it appeared that the presence of long tails in the microscopic disorder 
\begin{eqnarray}
P(V) \opsimeq_{  V  \to \pm \infty} 
\frac{1}{\vert V \vert^{1+\mu} }
\label{loimu}
\end{eqnarray}
was able to change the exponents \cite{zhangmu,marconi,krug}, 
as well as the 
morphology of the associated ultrametric tree structure of the locally optimal paths \cite{halpinmu}, even if the variance was finite $\mu>2$.

\subsection{ The strategy of finding the best site energy at zero-temperature}

For the disorder (\ref{loimu}),
the heuristic argument \cite{zhangmu,krug}
that has been proposed to explain these dependences in $\mu$ 
consists in
the balance between the maximal site energy $V_{max}=(L R^d)^{1/\mu}$ 
drawn in the volume $(L R^d)$
and the global elastic energy $R^2/L$, that yields the exponent
\begin{eqnarray}
\nu_{(Vmax)}^{elastic}(\mu)= \frac{\mu+1}{2 \mu-d}
\label{nuVmax}
\end{eqnarray}
This exponent corresponds to the strategy of finding the
maximal site energy and is usually considered as a lower bound 
 \cite{krug} for the true exponent $\nu(\mu)$.
For instance in $d=1$, this exponent is bigger than 
the value $\nu_{Gauss}=2/3$ for $\mu \ge 5$, which means that, 
at least for $\mu \ge 5$, the true exponent $\nu(\mu)$
is not the usual one $\nu_{Gauss}=2/3$. 
There has been a large number of numerical studies
on the true exponent $\nu(\mu)$, either on the directed polymer problem
or in corresponding growth models :
it is usually believed \cite{revue} that
the exponent (\ref{nuVmax}) is not the exact one \cite{zhangmu,krug} 
and in particular that the associated critical value $\mu_c(d=1)=5$ in $d=1$
is too low \cite{bourbonnais} and should be replaced by at least
$\mu_c(d=1)\sim 7$, even if it has been also argued
that these numerical conclusions were due to cross-over effects
and that the exponent (\ref{nuVmax}) was exact \cite{sander}.

From our point of view, it is not very clear why the exponent (\ref{nuVmax})
should describe the exponents at zero temperature of the
lattice models that are measured in the numerical simulations.
 Indeed, for lattice models at zero temperature, there is no elastic energy,
since a path is either allowed or forbidden.
It is thus useful to reconsider in this context the strategy of finding
the maximal site energy $V_{max}$ among a number of order
 $L^{1+d}$ of independent sites that can be reached for a path of length $L$.
The maximal value is thus of order $V_{max} \sim L^{(1+d)/\mu}$,
and since the distribution of its position is uniform,
it reads in terms of the longitudinal coordinate $l$ and the
transverse distance $r$ in continuum notations
\begin{eqnarray}
\rho_L(l,r) \sim \frac{ r^{d-1} \theta(r<l<L) }{ L^{d+1} } 
\end{eqnarray}
As a consequence, the probability distribution of the transverse distance $r$
of the best site $V_{max}$ takes the scaling form
\begin{eqnarray}
\rho_L(r) = \int_0^L dl \rho_L(l,r) 
\sim  \frac{1}{L} \left( \frac{r}{L} \right)^{d-1} \left( 1- \frac{r}{L} \right)
\end{eqnarray}
i.e. the exponent corresponding to the strategy of finding the best site
energy is 
\begin{eqnarray}
\nu_{(Vmax)}^{(lattice)}(\mu,d)= 1
\end{eqnarray}
and any deviation from this extremal value $\nu=1$ on the lattice
has to come from some cooperative effect of a large number
of subleading best sites. 
In the following, we will not discuss the zero-temperature anymore,
but the finite temperature case, and we will try to describe the cooperative 
effects by reconsidering our previous
 Imry-Ma arguments to see what changes
 are necessary in the presence of algebraic tails.
So we first need to consider the random energy that can be gained in a tube.

\subsection{ Distribution of the energy $U$
of a path of length $l$ in a tube of radius $r$ }

Let us consider the random energy 
\begin{eqnarray}
U = \sum_{i=1}^l V(i,x(i))
\end{eqnarray}
of a path of length $l$ in a tube of radius $r$ in dimensions $(1+d)$.
It is clear that its moments diverge for $k>\mu$
as the moments of the initial law (\ref{loimu}).
The tail of its probability distribution
 will thus present the same power-law decay
\begin{eqnarray}
P(U) \opsimeq_{ \vert U \vert \to \infty} \frac{ c(l,r) }{\vert U \vert^{1+\mu}}
\label{tailumu}
\end{eqnarray}
where the scaling of the prefactor $c(l,r)$ with respect to
$(l,r)$ can be estimated as follows. 
Let us introduce
temporarily a large cut-off $A$ in the original distribution (\ref{loimu})
to regularize the diverging moments $k>\mu$
\begin{eqnarray}
<V^k>_A \opsimeq_{A \to \infty} A^{k-\mu}
\end{eqnarray}
The largest divergence in the cut-off $A$ of the moment 
\begin{eqnarray}
<U^k>_A = \sum_{i_1} .. \sum_{i_k} < V(i_1,x(i_1)) ... V(i_k,x(i_k)) >_A
\label{momentsU}
\end{eqnarray}
is then given by the term of order $<V^k>_A$, 
corresponding to coincident indices $i_1=i_2=..=i_k$ and 
coinciding transversal positions $x_{i_1}=...=x_{i_k}$.
If one assumes a uniform confinement in transverse directions
within a tube of radius $r$, we thus obtain
\begin{eqnarray}
<U^k>_A \opsimeq_{A \to \infty} A^{k-\mu} \frac{l}{r^{d (k-1)}}
\end{eqnarray}
i.e. the prefactor $c(l,r)$ of the tail (\ref{tailumu}) scales as
\begin{eqnarray}
c(l,r) \sim \frac{ l }{ r^{d (\mu-1)}  }
\label{clr}
\end{eqnarray}
On the other hand, the largest contribution in $l$ of the moments 
(\ref{momentsU}) with even $k$ (odd moments vanish by symmetry)
comes from the term corresponding to $(k/2)$ pairs of coinciding indices,
leading to
\begin{eqnarray}
<U^k>_A \opsimeq_{l \to \infty} \left(<V^2> \frac{l}{r^d} \right)^{k/2}
\end{eqnarray}
The natural rescaling appropriate for this term
corresponds as it should to the rescaled variable $u$ 
introduced before for the Gaussian case
(\ref{imrymaarg}) 
\begin{eqnarray}
U= u \sqrt{ \frac{l}{r^d} }
\label{defsmallu}
\end{eqnarray}
However, the rescaled variable $u$ will now present the following algebraic tail
(\ref{tailumu},\ref{clr}) 
\begin{eqnarray}
{\cal P}(u) \opsimeq_{ \vert u \vert \to \infty} 
\frac{ (l r^d)^{1-\frac{\mu}{2} }  }
{\vert u\vert^{1+\mu}}
\label{tailsmallumu}
\end{eqnarray}
The presence of a non-trivial radius $r$ thus generalizes
the special case $r \sim 1$ corresponding to the classical problem
of the sum $U$ of $l$ independent variables :

(i) For $\mu>2$, the weight $(l r^d)^{1-\frac{\mu}{2}}$ of the tail of the rescaled variable
$u=U/\sqrt{l}$ vanishes in the limit
$l \to \infty$, i.e. there is a generalized Central Limit Theorem 
describing the limit law of the
rescaled variable $u$ as soon as the variance is finite.
However, since the polymer will try to find the best tube available,
the presence of the algebraic tail can induce a change in global exponents
as we will see.

(ii) For $\mu<2$, the weight of the tail
in (\ref{tailsmallumu}) is a now positive power of $l$ :
the variable $u$ is not appropriate anymore, and
the new appropriate rescaled variable $v$ is then defined by
(\ref{tailumu},\ref{clr}) 
\begin{eqnarray}
U= v  L^{\frac{1}{\mu} } R^{ d \frac{ 1- \mu }{ \mu } } 
\label{defsmallv}
\end{eqnarray}
so that the limit probability distribution of $v$
presents the tail
\begin{eqnarray}
Q(v) \opsimeq_{ \vert v \vert \to \infty} 
\frac{ 1 }
{\vert v \vert^{1+\mu}}
\label{tailvmu}
\end{eqnarray}
Since the exponent of the transverse scale $R$ in (\ref{defsmallv})
changes of sign at $\mu=1$, we will obtain a qualitative change at $\mu=1$.

\subsection{ Local Imry-Ma arguments}

In this Section, we discuss 
the `local' Imry-Ma arguments in the different cases $\mu>2$, $1<\mu<2$ and finally $0<\mu<1$.

\subsubsection{ Local Imry-Ma arguments for $\mu>2$ }

For $\mu>2$, the typical energy 
gained in a tube is still defined in terms
of the rescaled variable $u$ (\ref{defsmallu},\ref{tailsmallumu}),
and thus the exponents $(\omega_+,\nu_+)$
(\ref{R+versusL+},\ref{E+versusL+}) for typical favorable regions
and $(\omega_-,\nu_-)$
(\ref{R-versusL-},\ref{E-versusL-})
for typical unfavorable regions are unchanged.

\subsubsection{ Local Imry-Ma arguments for $1<\mu<2$ }

For $\mu<2$, the new appropriate rescaled variable is $v$ (\ref{defsmallv}),
and thus the typical energy associated to a tube is different from the Gaussian case. The Imry-Ma arguments for favorable and unfavorable regions
will thus yield different typical exponents.

(i) Analysis of the exponents in typical favorable regions

The free energy of a typical region
$(l_+,r_+)$ with the confinement term (\ref{defE+}) then becomes
\begin{eqnarray}
f_+  \sim T \frac{l_+}{r_+^2} 
- v _+ l_+^{\frac{1}{\mu} } r_+^{ d (\frac{1}{\mu}-1 )}
\label{defE+levy}
\end{eqnarray}
For a favorable region $v_+>0$, the minimization of $f_+$
with respect to $r_+$ 
 yields
\begin{eqnarray}
r_+ \sim v_+^{ - \frac{ \mu}{ (2-d) \mu +d} }  l_+^{\nu_+(\mu) }
\ \ {\rm with } \ \ \nu_+(\mu)= \frac{ \mu-1}{ (2-d) \mu + d}
\label{r+mu}
\end{eqnarray}
The corresponding free-energy reads
\begin{eqnarray}
f_+ \sim - v_+^{  \frac{ 2 \mu}{ (2-d) \mu +d} }  l_+^{\omega_+(\mu) }
\ \ {\rm with } \ \ \omega_+(\mu)= \frac{ 2+d -d \mu }{ (2-d) \mu + d}
\label{f+mu}
\end{eqnarray}

For $0<d<2$, this analysis is valid for $1<\mu<2$ :
the confinement exponent varies between $\nu_+(\mu \to 1) \to 0$ and 
its usual value $\nu_+(\mu \to 2) \to 1/(4-d)$, 
and the free-energy exponent varies between $\omega_+(\mu \to 1) \to 1$
 and its usual value $\omega_+(\mu \to 2) \to (2-d)/(4-d)$.
In particular, in dimension $d=1$, the exponents read
\begin{eqnarray}
\nu_+(\mu,d=1) && = \frac{ \mu-1}{  \mu + 1}
\\
 \omega_+(\mu,d=1) && = \frac{ 3 - \mu }{  \mu + 1}
\end{eqnarray}

For $d \geq 2$, these exponents are expected to be valid as long as 
the confinement exponent $\nu_+(\mu)$ is positive
and smaller than the free value $1/2$, i.e. for $\mu<1+2/d$
For instance, in transverse dimension $d=2$, the validity domain
is $1<\mu< 2$, whereas in $d=3$ it is $1<\mu< 5/3$.
Equivalently, for fixed $\mu$ in the interval $1<\mu<2$,
the exponents are valid for $d<2/(\mu-1)$.

(ii) Analysis of the exponents in typical unfavorable regions

The free energy of a typical region
$(l_-,r_-)$ with the elastic term (\ref{defE-}) then becomes
\begin{eqnarray}
f_-  \sim T \frac{r_-^2}{l_-} 
+ v _- l_-^{\frac{1}{\mu} } r_-^{ d (\frac{1}{\mu}-1 )}
\label{defE-levy}
\end{eqnarray}
For a unfavorable region $v_->0$, the minimization of $f_-$
with respect to $r_-$ 
 yields
\begin{eqnarray}
r_- \sim v_-^{  \frac{ \mu}{ (2+d) \mu -d} }  l_-^{\nu_-(\mu) }
\ \ {\rm with } \ \ \nu_-(\mu)= \frac{ \mu+1}{ (2+d) \mu - d}
\label{r-mu}
\end{eqnarray}
The corresponding free-energy reads
\begin{eqnarray}
f_- \sim v_-^{  \frac{ 2 \mu}{ (2+d) \mu -d} }  l_-^{\omega_-(\mu) }
\ \ {\rm with } \ \ \omega_-(\mu)= \frac{ 2+d -d \mu }{ (2+d) \mu - d}
\label{f-mu}
\end{eqnarray}

For $0<d<2$, this analysis is valid for $1<\mu<2$ :
the roughness exponent varies between $\nu_-(\mu \to 1) \to 1$ and 
its usual value $\nu_-(\mu \to 2) \to 3/(4+d)$, 
and the free-energy exponent varies between $\omega_-(\mu \to 1) \to 1$
 and its usual value $\omega_-(\mu \to 2) \to (2-d)/(4+d)$.
In particular, in dimension $d=1$, the exponents read
\begin{eqnarray}
\nu_-(\mu,d=1) && = \frac{ \mu+1}{ 3 \mu - 1} \\
 \omega_-(\mu,d=1) && = \frac{ 3 - \mu }{ 3 \mu - 1}
\end{eqnarray}

\subsubsection{ Local Imry-Ma argument for $0<\mu<1$ }

For $0<\mu<1$, the exponent of the transverse scale $R$ in (\ref{defsmallv})
is positive. As a consequence, exactly
 as in the case of long-range transverse correlation,
the favorable regions do not correspond to confined solutions, but 
to swollen solutions.  
The free-energy of a region $(l_-,r_-)$ 
with the elastic term (\ref{defE-}) has now to be replaced by
\begin{eqnarray}
f_-  \sim T \frac{r_-^2}{l_-} - v_-
L^{\frac{1}{\mu} } R^{ d (\frac{1}{\mu}-1 )}
\end{eqnarray}
For a favorable region $v_-<0$, the minimization with respect to $r_-$ yields
\begin{eqnarray}
r_- \sim v_-^{  \frac{ \mu}{ (2+d) \mu -d} }  l_-^{\nu_-(\mu) }
\ \ {\rm with } \ \ \nu_-(\mu)= \frac{ \mu+1}{ (2+d) \mu - d}
\label{r-mu1}
\end{eqnarray}
and the corresponding free-energy reads
\begin{eqnarray}
f_- \sim - v_-^{  \frac{ 2 \mu}{ (2+d) \mu -d} }  l_-^{\omega_-(\mu) }
\ \ {\rm with } \ \ \omega_-(\mu)= \frac{ 2+d -d \mu }{ (2+d) \mu - d}
\label{f-mu1}
\end{eqnarray}
In particular, in the limit $\mu \to 1$,
the free-energy exponent $\omega_-(\mu \to 1^-) =1$
is in continuity with the $\omega_+(\mu \to 1^+) =1$ (\ref{f+mu}).

\subsection{Global optimization for the full polymer  } 

\label{global3terms}

In the presence of long tails, the global optimization for the full polymer
consists in looking for the confinement scale
$R_S \sim L^{\nu_S(\mu)}$ and the global wandering scale
$R_G \sim L^{\nu(\mu)}$ that minimize the global free energy 
containing three contributions
\begin{eqnarray}
F_L(R_G,R_S)= T \frac{R_G^2}{L} + T \frac{L}{R_S^2} 
- U_{max}(L,R_S ; N=\frac{R_G^d}{R_S^d})
\label{freetotmu}
\end{eqnarray}
The first term $R_G^2/L $ represents the global elastic cost for the diffusive tube. The second term $\frac{L}{R_S^2}$ represents the 
confinement entropy cost. The third term $U_{max}$ represents the best energy
among $N$ independent variables $U$, where $U$ represents
the random energy associated to a tube $(L,R_S)$. Using the form  
(\ref{tailumu},\ref{clr}) of the tail of its probability distribution,
we thus obtain the following scaling
\begin{eqnarray} 
U_{max}(L,R_S ; N=\frac{R_G^d}{R_S^d}) 
\sim \left( c(L,R_S) N  \right)^{ \frac{1}{\mu}} 
\sim R_S^{-d} \left( L R_G^d \right)^{ \frac{1}{\mu}} 
\label{umaxrsrg}
\end{eqnarray}
This result shows why the introduction of two transverse scales $(R_S,R_G)$
is actually always better to gain energy than the
swollen solution that involves only one scale $R_G$,
whose free-energy reads
\begin{eqnarray}
F_L^{swollen}(R_G)= T \frac{R_G^2}{L}  
- U_{typ}(L,R_G )   
\end{eqnarray}
where the typical value scales as
 $U_{typ}(L,R_G) \sim R_G^{-d} \left( L R_G^d \right)^{ \frac{1}{\mu}}$ 
for $\mu<2$ (\ref{defsmallv}). The comparison with (\ref{umaxrsrg})
is thus immediate : both have the same factor 
$\left( L R_G^d \right)^{ \frac{1}{\mu}}$ presenting the scale of the best
sites in the volume $(L R_G^d)$, but the factor of the density in the transverse
directions of the tube is much better in a confined tube
$R_S^{-d}$ than for the swollen solution $R_G^{-d}$.

The optimization of (\ref{freetotmu}) with respect to $(R_S,R_G)$ yields
the following exponents
\begin{eqnarray}
\nu(\mu) && = 
\frac{ 1 + \mu (1-d) }{ (2-d) \mu -d }
\\
 \nu_S(\mu)  && = 
\frac{  \mu -(d+1)  }{ (2-d) \mu -d }
\label{soluglobal}
\end{eqnarray}

These exponents are valid in transverse dimension $0<d<2$ 
in the domain
\begin{eqnarray}
\mu_s(d)=d+1 < \mu < \mu_c(d)
\end{eqnarray}
where the inferior value $\mu_s(d)=d+1$ corresponds to
the point where the confinement exponent vanishes $\nu_S(d+1)=0$,
whereas the superior value corresponds the critical value 
 \begin{eqnarray}
 \mu_c(d) =  
 \frac{ 2( 1+d \nu) }{ (1+d \nu_+)} = 2 + d - \frac{d^2}{2} 
\label{mucriti} 
\end{eqnarray}
where the Gaussian exponents are recovered. 
This critical value varies between $\mu_c(d \to 0) \to 2$
and $\mu_c(d \to 2) \to 2$, and is maximal for $d=1$
with the value $\mu_c(d=1)=5/2$.
For instance, in dimension $d=1$, the exponents (\ref{soluglobal})  
 \begin{eqnarray}
\nu(\mu,d=1) &&= \frac{ 1 }{  \mu - 1  }
\\
\nu_S(\mu,d=1) && = \frac{ \mu- 2  }{  \mu - 1 }
\end{eqnarray}
are thus expected to be valid for $2 \leq \mu \leq \mu_c(d=1)=5/2$.

Let us now discuss what happens for $\mu<\mu_s(d)=d+1$.
Below this value, the confinement exponent would become
negative, which is unphysical, and thus the confinement exponent 
will stick to its minimal value 
\begin{eqnarray}
 \nu_S(\mu<\mu_s(d)=d+1)   = 0
\label{confinf}
\end{eqnarray}
corresponding to an extreme confinement. 
In this regime, the confinement entropy of order $L$ is subleading
 with respect to the elastic entropy and the disorder energy
in (\ref{freetotmu}). The balance between these two terms then coincides
with the estimation (\ref{nuVmax}) coming from the single best site strategy
\begin{eqnarray}
\nu( \frac{d}{2}  < \mu <\mu_s(d)=d+1) =
 \nu_{(Vmax)}^{elastic}(\mu)= \frac{\mu+1}{2 \mu-d}
\label{soluelas}
\end{eqnarray}
Note that this wandering exponent is then always greater than one.
As a consequence, on a lattice, the wandering exponent will stick
to its maximal value $\nu=1$
\begin{eqnarray}
\nu^{lattice}(   \mu \leq \mu_s(d)=d+1) =  1
\label{lattice}
\end{eqnarray}

\subsection{ Discussion }

In this Appendix, we have shown how the presence of algebraic tails
in the initial distribution could be taken into account in our approach
to yield a change in the global exponents even in the cases
where the variance is finite $\mu>2$. 
However, even if our results are on this point qualitatively correct,
the status of the quantitative results is not clear.
Indeed, in dimension $d=1$,
we have obtain as $\mu$ varies for the wandering exponent 
at non-zero temperature 
\begin{eqnarray}
\nu(d=1, 2< \mu < 5/2 ) && = \frac{ 1  }{  \mu -1 } \\
\nu(d=1, 1/2 < \mu <2) && = \frac{\mu+1}{2 \mu-1} \geq 1
\ \ { \rm i.e. } \ \ 
\nu^{lattice}(   \mu \leq 2) = 1 
\label{summd1} 
\end{eqnarray}
where the change that takes place at $\mu=2$ with
the corresponding value $\nu(\mu=2)=1$, between a phase
with a positive confinement exponent $\nu_s(\mu>2,d=1)>0$
and a phase with a vanishing confinement exponent $\nu_s(\mu<2,d=1)=0$.
The results (\ref{summd1}) for $\mu>2$
are far from the exponents measured at $T=0$
in numerical simulations : in particular, the critical value 
we have obtained $\mu_c(d=1)=5/2$ is very far
 from the lattice numerical estimation
$\mu_c(d=1,T = 0) \sim 7$ \cite{bourbonnais,revue}!
Is there a problem in this case between the zero-temperature best path
and our approach at non-zero temperature based on the confinement entropy?
Or is our Imry-Ma approach too simple to describe correctly
the algebraic tails ? For instance, it may be that the smaller scales
have to be taken into account, in contrast with the Gaussian
case where the optimization on the biggest scale fixes the exponents.

\end{document}